# Both Inter- and Intra-molecular Coupling of O-H Groups Determine the Vibrational Response of the Water / Air Interface

*Jan Schaefer, Ellen H. G. Backus, Yuki Nagata, and Mischa Bonn\**

*\*corresponding author: bonn@mpip-mainz.mpg.de*

Max Planck Institute for Polymer Research, Ackermannweg 10, 55128, Mainz, Germany

**Abstract:**

Vibrational coupling is relevant not only for dissipation of excess energy after chemical reactions but also for elucidating molecular structure and dynamics. It is particularly important for O-H stretch vibrational spectra of water, for which it is known that, in bulk, both intra- and intermolecular coupling alter the intensity and lineshape of the spectra. In contrast to bulk, the unified picture of the inter- / intra-molecular coupling of O-H groups at the water-air interface has been lacking. Here, combining sum-frequency generation experiments and simulation for isotopically diluted water and alcohols, we unveil effects of inter- and intra-molecular coupling on the vibrational spectra of interfacial water. Our results show that both inter- and intra-molecular coupling contribute to the O-H stretch vibrational response of the neat $H_2O$ surface – with intramolecular coupling generating a double-peak feature, while the intermolecular coupling induces a significant red-shift in the O-H stretch response.

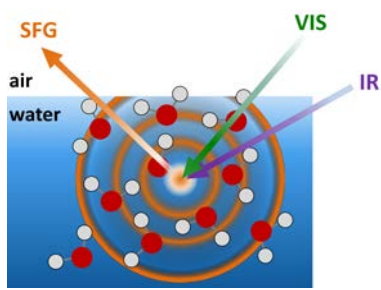

Vibrational spectroscopy techniques are widely used to obtain structural and dynamic insight into macroscopic ensembles.[1-2] Local vibrational modes are probed to extract molecular-level information about the structure and dynamics of liquids, such as the intermolecular interaction strengths between solute and solvent, and the binding sites of biomolecules.[3] However, the vibrational response of molecules often deviates from their vibrational density of states due to mode coupling, complicating the interpretation of the vibrational responses. Connecting the vibrational spectra to molecular-level information on the structures requires an understanding of these coupling effects. Vibrational coupling relies on interactions between different vibrations and requires a close proximity of the atoms involved. The coupling becomes strongest when the different modes are energetically (nearly-) degenerated. For instance, fundamental and overtone states of different modes with similar energy levels can be coupled with each other, leading to population of the (optically generally dark) overtone and an energy splitting, known as Fermi Resonance (FR). Moreover, coupling between different molecules of the same species can occur, giving rise to mode delocalization, often resulting in a broadened and / or enhanced response.[2]

The O-H stretch mode of condensed (H-bonded) water is a typical case for which coupling is known to critically affect the vibrational spectra. As has been shown theoretically[4-8] and inferred from time-resolved vibrational measurements,[9-11] vibrational mode coupling can occur between O-H stretch-oscillators of both the same and different water molecules and thus in an intra- and intermolecular fashion. These couplings enhance the speed of spectral diffusion through intermolecular vibrational energy transfer[10, 12] and vibrational energy transfer from the O-H stretch mode to the H-O-H bend mode.[13-16] This characterizes the unique vibrational dynamics of water such as a marked frequency dependence of the vibrational relaxation time,[17] and ultrafast anisotropy decay on a time scale of ~100 fs.[9-10, 18]

For interfacial water, the situation is different than for bulk water since the hydrogen bonding network is interrupted at the interface causing reorganization of water molecules. The structure of interfacial water has been studied using vibrational sum-frequency generation (V-SFG)-spectroscopy.[19-21] V-SFG is a surface-specific vibrational technique that provides the vibrational response of the outermost few water layers at an interface. This technique has revealed the presence of the dangling O-H group at the water / air interface[22] and similar hydrogen bond structure to that in the bulk.[17, 20-21, 23] Different reports exist on the effects of the intra- / inter-molecular couplings of water; so far, the experimentally determined response of the O-H stretch vibration at the water / air interface has been extensively discussed in terms of anharmonic FR coupling of intramolecular nature,[23-25] while the effects of intermolecular couplings on the O-H stretch SFG feature has been theoretically proposed.[26-29] A systematic discussion of, and a decomposition into, intramolecular and intermolecular couplings of interfacial water has not been reported, to the best of our knowledge.

In this work, we elucidate the role of inter- and intra-molecular vibrational couplings of the O-H stretch mode at the water / air, methanol / air and ethanol / air interfaces. Since ethanol and methanol cannot have intramolecular coupling between OH groups, the variation of the SFG spectra of these liquids and their isotopically diluted versions can be ascribed solely to intermolecular coupling. The SFG spectra of these alcohols show the impact of the intermolecular coupling of the O-H stretch mode on the SFG spectra, while the presence of a double-peaked feature in the response of the water / air SFG spectra indicates that the water O-H stretch spectra cannot be explained solely by intermolecular coupling. This suggests competing effects of the FR and intermolecular couplings. Molecular dynamics (MD) simulation support these experimental observations; the computed SFG spectra illustrate that the O-H stretch SFG feature is red-shifted due to the intermolecular couplings and reveals that a splitting of the energy level occurs due to the intramolecular coupling, which generates a double-peaked feature.

V-SFG spectroscopy was employed, combining a broadband femtosecond infrared laser pulse with a narrowband visible pulse. All spectra presented here were carried out in ssp polarization combination (*s*-polarized SFG, *s*-polarized visible and *p*-polarized IR) and at incident angles of 35° (SFG), 34° (VIS) and 36° (IR) with respect to the surface normal. Each spectrum results from averaging six 10-minute spectral accumulations, normalized by a non-resonant sum frequency signal from a *z*-cut quartz crystal and Fresnel-corrected using refractive indices reported elsewhere.[30-31] Details of the experimental SFG spectroscopy as well as the IR and Raman VV measurements are given in the Supporting Information. For the calculated SFG spectra based on the MD simulation, we ran a total 10 ns MD trajectories with the POLI2VS force field.[32] The 26.6 Å × 26.6 Å × 100 Å simulation cell contained 500 water molecules. The spectra were computed based on the dipole moment-polarizability correlation function,[33] within the truncating response function scheme.[26] The details of the simulations are also given in the Supporting Information.

First, we measured the V-SFG responses of C-H and O-H stretch modes of neat ethanol and methanol as well as the O-H stretch mode of neat $H_2O$. The spectra are displayed in Figure 1. The SFG spectral features are consistent with previous studies[34-38] (see the SI for a more detailed discussion). The V-SFG spectra result from a combined infrared-Raman transition. As such, to examine the spectral features of the interfacial molecules with those of the bulk molecules, we obtained the IR and Raman (VV) spectra ($I_{IR}$ and $I_{Raman(VV)}$, respectively) and constructed the IR*Raman(VV) spectra from $I_{IR} \times I_{Raman(VV)}$. The SFG spectral shapes of neat water, methanol, and ethanol are well-captured by the corresponding IR*Raman(VV) spectra. This indicates that the spectral features of the interfacial molecule are similar to those in the bulk. Furthermore, one can see that the C-H stretch mode peaks are much larger than those from O-H stretch modes. This causes the large interference of the C-H stretch peak of methanol and ethanol and the non-resonant background generating a tilted off-resonant contribution in the O-H stretch region, shown in Figure 2.

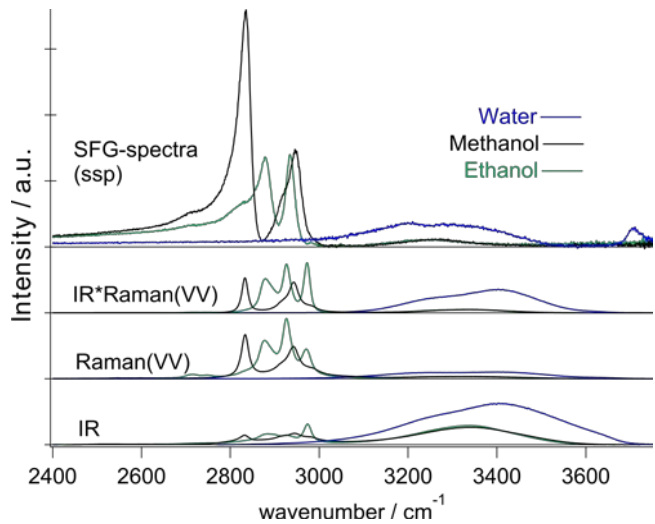

**Figure 1: The measured bulk IR, Raman(VV) and constructed IR*Raman(VV) spectra of water (blue), methanol (black) and ethanol (green) as well as the Fresnel-corrected SFG(ssp) $\left( = \left| \chi_{yyz}^{(2)} \right|^2 \right)$ spectra of corresponding liquid / air interfaces. For the SFG spectra, each tick represents ~1*10$^{-42}$ m$^4$/V$^2$, based on a bulk $\chi^{(2)}$ value for Quartz of 8.0*10$^{-13}$ m/V.[39]**

Figure 2 shows the O-H-stretch SFG spectra of isotopically diluted water, methanol and ethanol, where the O-H group of these molecules are (partly or completely) replaced with O-D groups. Comparison of the SFG spectra of neat water, methanol, and ethanol with those of isotopically diluted spectra provides information about the impact of intermolecular coupling on the vibrational response of these interfacial H-bonded structures. The underlying idea is that the variation of the O-H stretch SFG features for the neat and isotopically diluted methanol and ethanol can be uniquely linked to the intermolecular coupling of the O-H stretch mode, since there is no intramolecular OH-coupling, unlike for water. Therefore, we first focus on the SFG spectra of these alcohols.

The SFG spectra are depicted in Figure 2A and 2B. The disappearance of vibrational intensity on the red side of the O-H stretch peak upon isotopic dilution indicates that the O-H stretch modes of both alcohols are coupled intermolecularly with other O-H vibrational chromophores at the liquid / air interface; the intermolecular coupling of the O-H stretch mode broadens the O-H stretch peak and red-shifts the peak frequency. The variation of the SFG spectra closely resembles that of the constructed IR*Raman(VV) spectra (bottom panel in Figure 2), indicating that this coupling is commonly found both, at the interface and in the bulk.

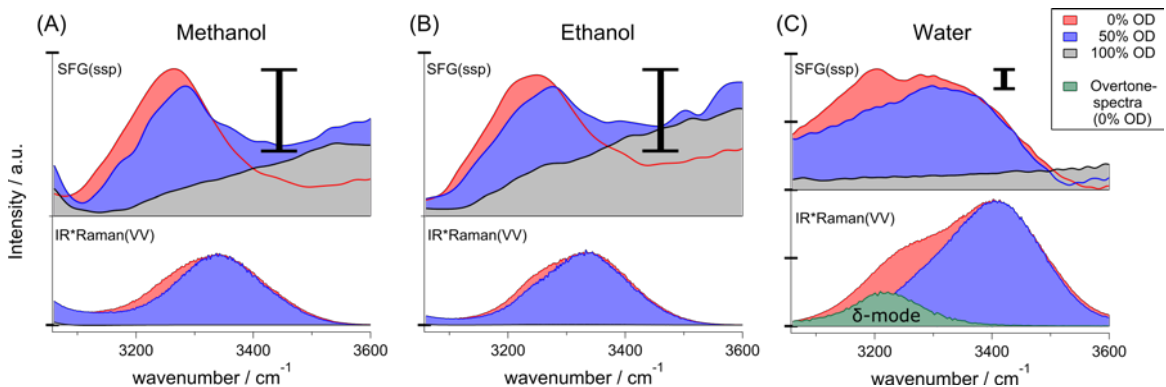

**Figure 2: Fresnel-corrected SFG (ssp)** $\left(=\left|\chi^{(2)}_{yyz}\right|^2\right)$ **and bulk IR*Raman(VV) spectra of Methanol (A), Ethanol (B) and Water (C) for 100% OH (red), 50% OH / 50% OD (blue) and 100% OD (black). For comparison, each 50% OH dilution spectrum is scaled up with its maximum to match the corresponding 100% OH spectrum. For the alcohol SFG spectra, the intensity scale is enlarged by a factor of 4, compared to that of water, as indicated by the scalebars. The slope in the background arising from interference between the large CH-resonances and the non-resonant contribution is explained in detail in the SI. For the constructed IR*Raman(VV) spectra of bulk water, the water bending mode overtone spectrum is also plotted with 4% anharmonicity[40] and multiplied in amplitude by a factor of 10. Original spectra and smoothing method are presented in the SI.**

Figure 2C shows corresponding spectra of neat, and isotopically diluted water. The double-peak feature of the hydrogen-bonded O-H stretch SFG spectra for neat $H_2O$ is merged into a single peak for the isotopically diluted $H_2O$. This trend is consistent with previous observations.[20, 41] Although this isotopic dilution approach itself cannot disentangle the influence of intra- and intermolecular couplings on the O-H stretch SFG spectra, the comparison of the SFG spectra variation of water with those of methanol / ethanol gives a hint for the effects of the intra- and inter-molecular couplings. The SFG spectra of all three liquids exhibit an enhanced SFG intensity around ~3200 cm$^{-1}$; this seems to originate from intermolecular interactions of O-H groups. For water, moreover a double-peak feature is apparent, which cannot be observed in the SFG spectra of methanol or ethanol. So far, this double-peak feature in the water SFG spectrum has been ascribed to the anharmonic FR coupling of the O-H stretch mode and H-O-H bending overtone.[25] The response of the H-O-H bending overtone in bulk water is also depicted in Figure 2C (bottom panel).

To experimentally explore whether the anharmonic FR can generate such a double-peak feature of the O-H stretch mode, we compare the O-H stretch features with the O-D stretch features of neat methanol and isotopically diluted methanol in the constructed IR*Raman(VV) spectra. The spectra are presented in Figure 3. For methanol, the O-H stretch region is decoupled from the overtone of the H-O-C bending mode. Therefore, the variation of the O-H stretch spectra must arise solely from the intermolecular coupling, as is illustrated in Figure 2A. In contrast, the overtone of the bending mode overlaps with the O-D stretch band (2200-2600 cm$^{-1}$), which results in a FR. As such, the variation of the O-D stretch spectra of deuterated methanol will be affected by both the FR and intermolecular coupling. Remarkably, while the O-H stretch spectrum looks smooth, the O-D stretch spectrum of methanol exhibits a double-peak feature, much like that observed for the pure water spectrum. This demonstrates that the FR is a very likely candidate to explain the double-peak feature of the SFG spectra of water.

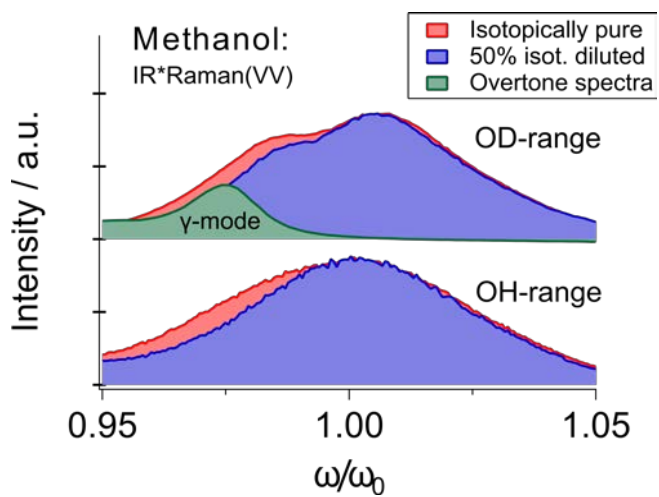

**Figure 3: Constructed IR*Raman(VV) spectra of methanol within the OH- and OD-stretching region, normalized to their central frequencies and scaled to equal intensity. For both ranges, the isotopically pure (100% OH or OD, red), the 50% dilution (blue) and the overtone spectrum ($\omega_{ov}=\omega_{fundamental}$*1.96) are plotted. The overtone spectrum is multiplied by a factor of 10.**

Above, our experimental observations show that the intermolecular coupling between O-H stretch modes red-shifts the O-H stretch SFG peak, while the FR with the bending mode overtone can generate the double-peak feature in the SFG spectra of neat H$_2$O at the water / air interface. To verify these conclusions, we performed MD simulations at the neat H$_2$O / air and neat HDO / air interface and computed the O-H stretch SFG spectra of H$_2$O and HDO. The SFG spectrum calculated with the autocorrelation (AC) term for neat HDO (HDO (AC) spectrum) corresponds to the spectrum of an isolated O-H group in $D_2O$. The SFG spectrum with the cross correlation (CC) term for neat $H_2O$ corresponds to the neat $H_2O$ SFG spectrum ($H_2O$ (AC+CC) spectrum). The difference between the spectra calculated with the autocorrelation (AC) terms of neat $H_2O$ ($H_2O$ (AC) spectrum) and the $H_2O$ (AC+CC) spectrum is the contribution of the intermolecular O-H stretch mode couplings. The difference between the $H_2O$ (AC) and HDO (AC) spectra can be considered as the contribution of intramolecular coupling, including the FR contribution, to the overall response. The imaginary part of the SFG response for HDO (AC), $H_2O$ (AC), and $H_2O$ (AC+CC) are shown in Figure 4A.

The difference between the HDO (AC) and $H_2O$ (AC) spectra indicates the intramolecular couplings including the FR can enhance the spectral amplitude at 3300 cm$^{-1}$ and 3440 cm$^{-1}$, causing a vibrational energy splitting by 140 cm$^{-1}$: the intramolecular coupling of the O-H stretch chromophores does not uniformly enhance the H-bonded O-H stretch band. This illustrates that the intramolecular coupling causes energy splitting, generating the double-peak feature, consistent with the experimental observation. However, the difference between the $H_2O$ (AC) and HDO (AC) spectra is relatively small compared with the difference between the $H_2O$ (AC+CC) and $H_2O$ (AC) spectra. This indicates that the intermolecular coupling strongly affect the SFG spectra of water and red-shift the H-bonded O-H stretch band.

For comparison, experimentally determined $\mathrm{Im}(\chi^{(2)})$ spectra[42] of neat $H_2O$ and that of HDO are presented in panel B of Figure 4. The HOD response was obtained from the response of a 3:1 $D_2O$:$H_2O$ mixture, correcting for the statistically present $H_2O$ molecules in such a mixture[20, 41]. These $\mathrm{Im}(\chi^{(2)})$ spectra spectra are slightly different from literature spectra[20,41], which is probably caused by uncertainties in correcting for the phase in the heterodyne experiments[43,44]. Qualitatively, the experimental spectra agree with the simulated ones, also showing a decrease of intensity on the low frequency side of the absorption band upon isotopic dilution associated with the coupling mechanisms discussed above.

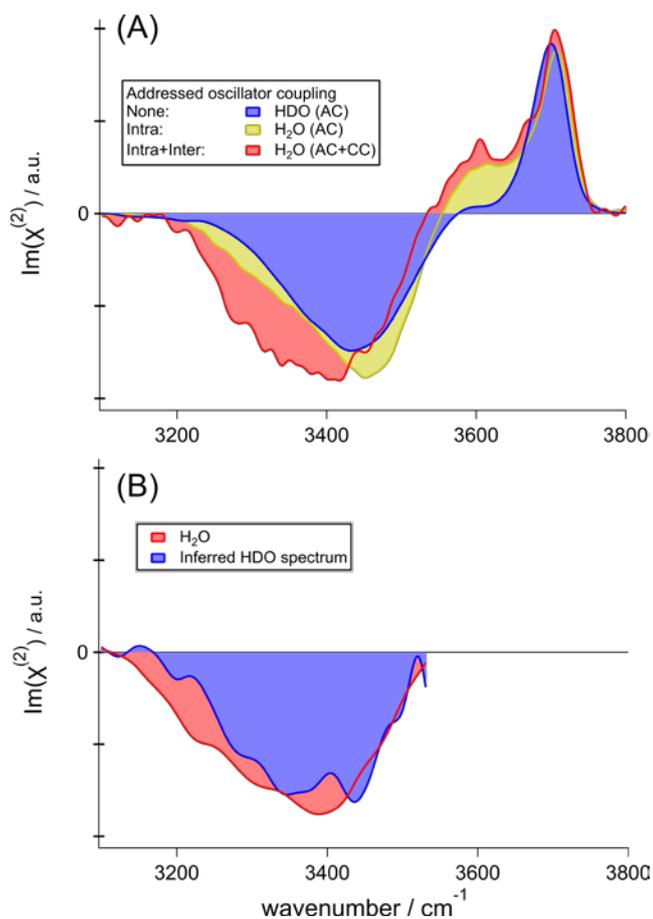

**Figure 4**: Imaginary part of $\chi^{(2)}$ for the water / air interface. (A): Spectra derived from AIMD simulations of HDO and $H_2O$. To address inter- and intra-molecular coupling separately, auto- (AC) and cross- (CC) correlation terms are calculated for both systems. (B): Measured response $(\text{Im}(\chi^{(2)}_{yyz}))$ of neat $H_2O$ and that of HDO, which is calculated by correcting the response from a $D_2O$:$H_2O$ mixture in the ratio 3:1, for the contribution of remaining $H_2O$ molecules .[42]

In conclusion, our results illustrate that, at the surface as in the bulk, intermolecular coupling results in red-shifted and broadened absorption bands of H-bonded O-H stretches. For O-H groups of short-chain alcohols, we find experimentally that there is a substantial contribution of intermolecular coupling to the vibrational SFG response of H-bonded O-H groups at the surface, in a similar fashion to responses in the bulk. A comparison between O-H- and O-D-bands revealed that anharmonic FR coupling results in energy splitting of these broad absorption bands while the major broadening effect is of intermolecular nature. MD simulations of the water / air interface agree with our experimental findings and further demonstrate that for the broadening of the H-bonded O-H stretching band, the difference spectra of neat $H_2O$ and isotopically diluted water arises more from inter- rather than intra-molecular couplings. Our results provide experimental evidence for the impact of the delocalized character in H-bonded O-H structures on their vibrational responses.

# Acknowledgments

We thank Saman Hosseinpour and Taisuke Hasegawa for fruitful discussions as well as Sapun Parekh and Jenée Cyran for a careful reading of the manuscript. This work was funded by an ERC Starting Grant (Grant No. 336679).